\documentstyle[pre,aps,preprint,epsfig]{revtex}

\begin{document}

\draft
\tighten

\preprint{\vbox{\hfill nlin.CD/0108024 \\
          \vbox{\hfill August 2001} \\
          \vbox{\vskip0.5in}
         }}

\title{The Approach to Ergodicity in the Quantum Baker's Map}

\author{
Andrew Jordan\footnote{E--mail: \tt ajordan@physics.ucsb.edu}
   and 
Mark Srednicki\footnote{E--mail: \tt mark@physics.ucsb.edu}
       }
\address{Department of Physics, University of California,
         Santa Barbara, CA 93106 
         \\ \vskip0.5in}

\maketitle

\begin{abstract}
\normalsize{
We study the quantum mechanics of a generalized version of the baker's map.
We show that the Ruelle resonances (which govern the approach to ergodicity of 
classical distributions on phase space) also appear in the quantum 
correlation functions of observables at different times, and hence
control the statistical variance of matrix elements of observables
(in the basis of eigenstates of the quantum time evolution operator).
We illustrate this with numerical results.
}
\end{abstract}


\section{Introduction}
\label{1}

The baker's map and its generalizations \cite{b,kh} provide prototypical 
models of chaotic systems, and thus their properties, both classical and 
quantum, have been extensively studied.  The issue we examine here is the
structure of the Ruelle resonances \cite{r} (which control the approach to 
ergodicity of smooth distributions on the classical phase space), 
and the role these resonances play in the properties of the corresponding
quantum map.

This question has been partly addressed in earlier work.  
In ref.\cite{hs}, the Ruelle resonances of the original baker's map
were computed, and the classical dynamics of phase-space distributions
were thoroughly investigated.
In ref.\cite{a}, the statistical distribution of the eigenvalues
of the quantum time-evolution operator $U$ for a generic chaotic system
was studied, and it was found that the Ruelle resonances control the 
deviations of this distribution from the Wigner-Dyson form that is
predicted by the random-matrix analogy \cite{rm}.
In ref.\cite{fish}, certain matrix elements of the resolvent of $U$
were studied (for the original baker's map), and used to identify 
particular Ruelle resonances.  Our work here is closely related, 
though we require less numerical effort to see the resonances in 
the quantum time-correlation functions that we examine.
We also uncover the role of the resonances in the statistical variance
of matrix elements of observables (in the eigenbasis of $U$).

We review the needed information about the classical map
in Sec.~\ref{2}, and about the quantum map in Sec.~\ref{3}.
In Sec.~\ref{4} we introduce the quantum correlation functions
that we study numerically in Sec.~\ref{5}.  Sec.~\ref{6}
contains our conclusions.  In an Appendix, we calculate the
values of the Ruelle resonances by computing the associated
spectral determinant or dynamical zeta function.

\section{Properties of the classical map}
\label{2}

Classically, the generalized baker's map 
$\cal M$: $x\to x'={\cal M}(x)$ \cite{b,kh}
is defined on a unit square (interpreted physically as a 
two-dimensional
phase space) $x=(q,p)$, $0\le q\le 1$, $0\le p\le 1$.
To implement the map, this square is first cut into $s$ vertical strips with
widths $w_1,\ldots,w_s$; $\sum_{a=1}^s w_a=1$.
Let $e_a=\sum_{b=1}^{a-1} w_b$ 
be the left edge of the $a^{\rm th}$ strip (with $e_1=0$ and $e_{s+1}=1$).
Each strip is then stretched horizontally and compressed vertically to 
unit width and height $w_a$, and the strips are stacked vertically to
assemble a new unit square.  We therefore have
\begin{equation}
\pmatrix{q' \cr \noalign{\smallskip} p' \cr} = \sum_{a=1}^s 
\pmatrix{(q-e_a)/w_a \cr \noalign{\smallskip} w_a p + e_a \cr}
\theta(e_a\leq q<e_{a+1}),
\label{baker}
\end{equation}
where we have introduced a generalized step function: 
\begin{equation}
\theta(S) = \cases{ 1 & if $S$ is true, \cr
                    0 & if $S$ is false. \cr }
\label{ths}
\end{equation}
Note that if $q$ is in the range of the $a^{\rm th}$ strip 
($e_a\leq q<e_{a+1}$), then so is $p'$ ($e_a\leq p'<e_{a+1}$).  
This point is important in the construction of the quantum map.

It is convenient to introduce a Dirac notation:  let
$|x)$ be a phase-space eigenstate with normalization
$(x|x') = \delta(x-x')=\delta(q-q')\delta(p-p')$.
(To avoid confusion we use parentheses for classical states
and angle brackets for quantum states.)
We define the Perron-Frobenius time evolution operator $\cal U$ via
\begin{equation}
(x|\,{\cal U}|x') \equiv \delta(x-{\cal M}(x')).
\label{ufp}
\end{equation}
If we consider the domain of $\cal U$ to be the
space of Lebesgue square-integrable functions on the unit square, 
then $\cal U$ is unitary for any area-preserving map $\cal M$, 
and so its spectrum lies on the unit circle.
However, if the domain of $\cal U$ is restricted to a suitable class of 
smooth 
(infinitely differentiable) functions, 
then $\cal U$ is effectively truncated, and it is no longer unitary in
the reduced space.  If this truncated $\cal U$ is put into Jordan 
(upper-triangular) form, the diagonal entries are inside the unit circle;
these diagonal entries are the Ruelle resonances $u_\ell$ of the map.  
Their values control the decay of initially smooth distributions on phase space
to the ergodic (uniform) distribution.

Hasegawa and Saphir \cite{hs} have shown that the Ruelle resonances
of the original baker's map ($s=2$, $w_1=w_2={1\over2}$) 
are given by $u_\ell = 2^{-\ell}$ with degeneracy $d_\ell = \ell+1$
for $\ell=0,1,2,\ldots\,$. 
Below, we show that for the generalized baker's map,
$u_\ell = \sum_{a=1}^s w_a^{\ell+1}$, with the same degeneracy.  
In general, the values of the
Ruelle resonances have no direct connection with the Lyapunov exponents
that govern the short-time instabilities of individual trajectories.  
For example, the positive Lyapunov exponent for a generic (nonperiodic) 
trajectory is $\lambda = -\sum_{a=1}^sw_a\ln w_a$.

Consider the time evolution of an initial smooth 
distribution on phase space, $A(x)=(x|A)$.  We have
\begin{eqnarray}
(x|\,{\cal U}^t|A)
&=& \int dx'\,(x|\,{\cal U}^t|x')(x'|A)
\nonumber \\
&=& \int dx'\,\delta(x-{\cal M}^t(x'))A(x')
\nonumber \\
&=& A({\cal M}^{-t}(x)).
\label{xua}
\end{eqnarray}
Here ${\cal M}^t$ denotes $t$ iterations of $\cal M$, and ${\cal M}^{-t}$
denotes $t$ iterations of the inverse map ${\cal M}^{-1}$; the last equality
in eq.~(\ref{xua}) follows from 
the change of variable $x'={\cal M}^{-t}(y')$, which
has unit jacobian if $\cal M$ is area preserving.

Let us now consider time correlation functions of the form 
$(B|\,{\cal U}^t|A)$,
where $A(x)=(x|A)$ and $B(x)=(x|B)$ are both smooth functions, 
and $t$ is an integer (positive or negative).  We have
\begin{eqnarray}
(B|\,{\cal U}^t|A)
&=& \int dx\,(B|x)(x|\,{\cal U}^t|A)
\nonumber \\
&=& \int dx\,B(x)A({\cal M}^{-t}(x))
\nonumber \\
&=& \int dx\,B({\cal M}^t(x))A(x).
\label{bua}
\end{eqnarray}
For the generalized baker's maps, these correlation functions can be written,
for $t\ge0$, as
\begin{equation}
(B|\,{\cal U}^t|A) = \sum_{\ell=0}^\infty
%                               c_\ell \, e^{-\gamma_\ell t}.
                                c_\ell(t)u_\ell^t ,
\label{ruelle}
\end{equation}
where $c_\ell(t)$ is a polynomial in $t$ of maximal order $d_\ell-1=\ell$.
The coefficient $c_\ell(t)$ depends on the choice of $A$ and $B$,
but the Ruelle resonance $u_\ell$ does not.  We always have
$u_0=1$ and $|u_\ell|<1$ for $\ell\ge 1$, and so initial correlations
decay as time evolves.  If we consider the case $A=B$, then eq.~(\ref{bua})
implies that we have time symmetry, and so
\begin{equation}
(A|\,{\cal U}^t|A) = \sum_{\ell=0}^\infty
%                               c_\ell \, e^{-\gamma_\ell |t|},
                                c_\ell(|t|)u_\ell^{|t|},
\label{ruelle2}
\end{equation}
where $c_\ell(|t|)$ is again a polynomial of maximal order $\ell$ in $|t|$.

It is useful to introduce a set of basis polynomials 
(actually shifted and rescaled Legendre polynomials) \cite{hs}
\begin{equation}
P_\ell(q) = {{\sqrt{2\ell+1}} \over \ell!}\,{d^\ell\over dq^\ell} 
\Bigl(q^\ell(1-q)^\ell\Bigr)
\label{pl}
\end{equation}
that are orthonormal on the unit interval,
$\int_0^1 dq\,P_{\ell'}(q)P_\ell(q)=\delta_{\ell'\ell}$.
We define a set of classical states $|\ell m)$ on phase space via
\begin{equation}
(x|\ell m) = P_\ell(q) P_m(p).
\label{xlm}
\end{equation}
According to eq.~(\ref{xua}), the action of the Perron-Frobenius operator  
on these states is given by
\begin{equation}
(x|\,{\cal U}|\ell m)
= \sum_{a=1}^s P_\ell(w_a q+e_a) P_m((p-e_a)/w_a) 
               \,\theta(e_a\le p < e_{a+1}).
\label{xulm}
\end{equation}
We wish to find the matrix elements of $\cal U$ in the $|\ell m)$ basis,
\begin{equation}
(\ell'm'|\,{\cal U}|\ell m) 
= \int dx\,P_{\ell'}(q)P_{m'}(p)(x|\,{\cal U}|\ell m).
\label{lpmpulm}
\end{equation}
In order to do so, we must evaluate integrals of the form 
$I_{\ell'\ell}=\int_0^1 dq\,P_{\ell'}(q)P_\ell(wq+e)$,
where $0 < w \le 1$ and $0 \le e \le 1-w$ are constants.
(The $p$ integrals can be put into this form by a change of variable
that results in an extra factor of $w$.)
Using eq.~(\ref{pl}) for $P_{\ell'}(q)$ and repeatedly integrating by parts,
we see that $I_{\ell'\ell}=0$ if $\ell'>\ell$, and it is similarly
straightforward to show that $I_{\ell\ell}=w^\ell$.  Thus the matrix 
$(\ell'm'|\,{\cal U}|\ell m)$ is in Jordan form (upper-triangular with
respect to the $\ell$ indices, and lower-triangular with respect to the
$m$ indices).  The diagonal elements are the Ruelle resonances, given by 
\begin{equation}
(\ell m|\,{\cal U}|\ell m) = \sum_{a=1}^s w^{\ell+m+1}_a \equiv u_{\ell+m},
\label{lmulm}
\end{equation}
in agreement with eq.~(\ref{uell}).  We also see that $u_\ell$ has degeneracy
$d_\ell=\ell+1$, since there are $\ell+1$ different $|\ell m)$ states that
result in the same value of $(\ell m|\,{\cal U}|\ell m)$.

Eq.~(\ref{lmulm}) also holds if we replace $\cal U$ with ${\cal U}^{-1}$
on the left-hand side.  Furthermore, because of the Jordan form of $\cal U$
in this basis, we have
\begin{equation}
(\ell m|\,{\cal U}^t|\ell m) = u_{\ell+m}^{|t|}
\label{lmutlm}
\end{equation}
for any integer $t$ (positive or negative). 
Thus the time-evolution of one of the $|\ell m)$  states picks out a 
particular Ruelle resonance.

\section{Properties of the quantum map}
\label{3}

To quantize the generalized baker's map, we must discretize 
$q$ and $p$ \cite{b};
let $q_j = (j-\nu_1)/N$ and $p_j = (j-\nu_2)/N$, where $j=1,\ldots,N$;
the integer $N$ plays the role of the inverse of Planck's constant $h$,
while $0 \le \nu_{1,2} \le 1$ are parameters of the discretization 
that should be irrelevant in the classical limit.  
Also, we require that the strip widths $w_a$ be integers divided by $N$.
We then define corresponding quantum states $|q_j\rangle$
and $|p_j\rangle$ with the properties
\begin{equation}
\langle q_j|q_k\rangle = \langle p_j|p_k\rangle = \delta_{jk}, 
\label{qq}
\end{equation}
\begin{equation}
\langle p_j|q_k\rangle = N^{-1/2}\exp(-2\pi i p_j q_k/N) \equiv 
(F_N)_{jk}.
\label{qp} 
\end{equation}
We now wish to specify the unitary time evolution operator $U$ for 
a single iteration of the map.  
From eq.~(\ref{baker}), we expect that, heuristically, 
$U|q_k\rangle \sim |(q_k-e_a)/w_a\rangle$
when $q_k$ is in the $a^{\rm th}$ strip. 
This must be reconciled, however, with the evolution of the momentum $p$.
Recall that if the initial coordinate $q$ is in the $a^{\rm th}$ strip,
then so is the final momentum $p'$.  This motivates a specification of $U$
in a mixed $p$-$q$ basis,
\begin{equation}
\langle p_j|U|q_k\rangle = 
\sum_{a=1}^s w_a^{-1/2}e^{i\chi_a}\langle p_j|(q_k-e_a)/w_a\rangle 
             \,\theta(e_a\le q_k<e_{a+1})
             \,\theta(e_a\le p_j<e_{a+1}).
\label{u}
\end{equation}
The prefactors of $w_a^{-1/2}$ are needed for unitarity, while the 
phase angles $\chi_a$ can be arbitrary linear functions of $q_j$ and $p_k$;
these phases should be irrelevant in the classical limit, since they
can be absorbed by shifts in the origins of $q$ and $p$ in the definition
of the inner product (\ref{qp}).  It is then convenient 
(and by now traditional) to choose these phases so that $U$ is given in the
coordinate basis by
\begin{equation}
\langle q_j|U|q_k\rangle = \Biggl(F_N^{-1}\Biggr)_{ji}
\pmatrix{
e^{i\phi_1} F_{Nw_1} & 0 & \ldots & 0 \cr
\noalign{\medskip}
0 & e^{i\phi_2} F_{Nw_2} & \ldots & 0 \cr
\noalign{\medskip}
\vdots & \vdots & \ddots & \vdots \cr
\noalign{\medskip}
0 & 0 & \ldots & e^{i\phi_s} F_{Nw_s} \cr}_{ik},
\end{equation}
where the $\phi_a$'s are arbitrary {\em constant\/} angles. 

Since $U$ is unitary, its eigenvalues are phases,
\begin{equation}
U|\alpha\rangle = e^{-i\theta_\alpha}|\alpha\rangle,
\label{ua}
\end{equation}
and the corresponding eigenstates are orthonormal,
$\langle\alpha|\beta\rangle = \delta_{\alpha\beta}$.

\section{Quantum time correlation functions}
\label{4}

Consider a quantum operator ${\cal O}(q,p)$ that is a smooth function
of the position and momentum operators $q$ and $p$, and that does not
depend explicitly on $\hbar = 1/2\pi N$.  Then Shnirelman's theorem \cite{shn}
states that the diagonal matrix element $\langle\alpha|{\cal O}|\alpha\rangle$,
where $|\alpha\rangle$ is an eigenstate of $U$, tends to the phase-space
average $\int dx\,{\cal O}(x)$ as $\hbar\to0$.  (For a proof in the case
of the baker's map, and a complete set of references, see \cite{shn2}.)
For hamiltonian systems, this result was generalized by 
Feingold and Peres \cite{fp} and by Wilkinson \cite{w} 
(for rigorous results see \cite{fpwr}) 
to the case ${\cal O}=B U^t\!A U^{-t}$, where $A(q,p)$ and $B(q,p)$ are
smooth, $N$-independent functions. 
In the Heisenberg picture, $U^t \!A(x) U^{-t}=A(x_t)$, where $x_t$ denotes
the time-evolved position and momentum operators.  If, however, we
use classical evolution instead, then according to eq.~(\ref{xua})
$x_t \to  {\cal M}^{-t}(x)$.  This substitution
should be valid in the $N\to\infty$ limit, and so
\begin{equation}
\langle\alpha| B U^t\!A U^{-t}|\alpha\rangle 
= \int dx\,B(x)A({\cal M}^{-t}(x)) + O(N^{-1/2}),
\label{fpw}
\end{equation}
where the $O(N^{-1/2})$ corrections can be expressed as a sum over periodic 
orbits \cite{w,fpwr}.  

If we insert a complete set of $U$ eigenstates on the left-hand side
of eq.~(\ref{fpw}), we have
\begin{eqnarray}
\langle\alpha| B U^t\!A U^{-t}|\alpha\rangle 
&=& \sum_\beta \langle\alpha|B U^t |\beta \rangle
               \langle\beta |A U^{-t}|\alpha\rangle
\nonumber \\
&=& \sum_\beta B_{\alpha\beta} A_{\beta\alpha} \, 
               e^{i(\theta_\alpha-\theta_\beta)t},
\label{fpw2}
\end{eqnarray}
where $B_{\alpha\beta}\equiv\langle\alpha|B|\beta\rangle$, etc.
Thus we get the Feingold-Peres-Wilkinson formula 
(applied to a quantum map rather than a hamiltonian system),
\begin{equation}
\sum_\beta B_{\alpha\beta} A_{\beta\alpha} \, 
           e^{i(\theta_\alpha-\theta_\beta)t}
= \int dx\,B(x)A({\cal M}^{-t}(x)) + O(N^{-1/2}).
\label{fpw3}
\end{equation}
Now, according to eq.~(\ref{bua}), the leading term on the right-hand side
of eq.~(\ref{fpw3}) is given by the right-hand side of eq.~(\ref{ruelle}).
Thus we have, for $t\ge 0$,
\begin{equation}
\sum_\beta B_{\alpha\beta} A_{\beta\alpha} \, 
           e^{i(\theta_\alpha-\theta_\beta)t}
= \sum_{\ell=0}^\infty c_\ell(t) e^{-\gamma_\ell t} + O(N^{-1/2})
\label{fpw4}
\end{equation}
where we have defined $e^{-\gamma_\ell} \equiv u_\ell$.  
This is our main result,
which we will investigate numerically in the next section. 

Also, since the leading term on the right-hand side of eq.~(\ref{fpw4})
is independent of $|\alpha\rangle$, we can improve the accuracy (by a factor
of $N^{-1/2}$) by averaging over $\alpha$,
\begin{equation}
{1\over N} \sum_{\alpha\beta} B_{\alpha\beta} A_{\beta\alpha} \, 
                                  e^{i(\theta_\alpha-\theta_\beta)t}
= \sum_{\ell=0}^\infty c_\ell(t) e^{-\gamma_\ell t} + O(N^{-1}).
\label{fpw5}
\end{equation}
Note also that the left-hand side of eq.~(\ref{fpw5}) can be expressed as
the quantum trace $(1/N)\mathop{\rm Tr}B U^t\!A U^{-t}$.

Let us consider the special case $A=B$; we then have
\begin{equation}
{1\over N} \sum_{\alpha\beta} |A_{\alpha\beta}|^2 \, 
                       e^{i(\theta_\alpha-\theta_\beta)t}
       = \sum_{\ell=0}^\infty c_\ell(|t|) e^{-\gamma_\ell |t|} + O(N^{-1}).
\label{fpw6}
\end{equation}
We see that the statistical variance of the matrix elements 
$A_{\alpha\beta}$ is controlled by the Ruelle resonances.  For example,
let us choose $A(x)=P_\ell(q)$; then a single Ruelle resonance appears
on the right-hand side of eq.~(\ref{fpw6}), and furthermore $c_\ell(|t|)=1$
in this case.  Thus we have
\begin{equation}
{1\over N} \sum_{\alpha\beta} |P_\ell(q)_{\alpha\beta}|^2 \, 
                       e^{i(\theta_\alpha-\theta_\beta)t}
= e^{-\gamma_\ell |t|} + O(N^{-1}).
\label{fpw7}
\end{equation}
We next multiply both sides by $e^{-i\omega t}e^{-\varepsilon|t|}$,
where $0<\varepsilon\ll\min(1,\gamma_\ell)$ provides a cutoff,
and sum over $t$; we get
\begin{equation}
{1\over N} \sum_{\alpha\beta} |P_\ell(q)_{\alpha\beta}|^2 
           \, {\varepsilon \over 
             \sin^2((\theta_\alpha-\theta_\beta - \omega)/2)
              + (\varepsilon/2)^2 }
= {\sinh \gamma_\ell \over \sin^2(\omega/2) + \sinh^2(\gamma_\ell/2) }
  +O(N^{-1}).
\label{fpw8}
\end{equation}
If we now take the formal limit of $\varepsilon\to 0$ on the 
left-hand side, the $\varepsilon$-dependent factor becomes
$2\pi\delta(\theta_\alpha-\theta_\beta - \omega)$.  This 
shows us that the statistical distribution of the matrix elements
$P_\ell(q)_{\alpha\beta}$ with fixed 
$\omega=\theta_\alpha-\theta_\beta$ is a Lorentzian in $\sin(\omega/2)$
whose width is $2\sinh(\gamma_\ell/2)$.  However, this formal limit
must be interpreted with care, and so in our numerical work
we concentrate on eq.~(\ref{fpw7}).

\section{Numerical Results}
\label{5}

We consider the example of a two-strip map with $w_1$ close to $1/3$
and $w_2$ close to $2/3$.  We then numerically evaluate
\begin{equation}
C_{\ell,\alpha}(t) \equiv 
\langle\alpha|P_\ell(q) U^t P_\ell(q) U^{-t}|\alpha\rangle
= \sum_\beta |P_\ell(q)_{\alpha\beta}|^2 \,
                       e^{i(\theta_\alpha-\theta_\beta)t}
\label{c}
\end{equation}
and its average
\begin{equation}
\overline C_\ell(t) \equiv 
{1\over N} \mathop{\rm Tr} P_\ell(q) U^t P_\ell(q) U^{-t}
= {1\over N} \sum_{\alpha\beta} |P_\ell(q)_{\alpha\beta}|^2 \, 
                       e^{i(\theta_\alpha-\theta_\beta)t}
\label{c2}
\end{equation}
for different values of $\ell$ and $N$.  We note that the numerical
problem of computing $\overline C_\ell(t)$ is considerably simpler
than that of computing $C_{\ell,\alpha}(t)$; to compute the latter,
we must diagonalize $U$, whereas the former can be done by taking a 
trace of a simple matrix product in the position representation.
We take the irrelevant parameters to be 
$\nu_1=\nu_2={1\over2}$ and $\phi_1=\phi_2=0$. 
We compare these results with the classical expectation 
$e^{-\gamma_\ell |t|}$ in fig.~(1) for $N=331$, 
the highest value we used; good agreement can be seen.

Because of the $O(N^{-1})$ corrections in eq.~(\ref{fpw6}), agreement
between $\overline C_\ell(t)$ and $e^{-\gamma_\ell |t|}$ breaks down
at later times.  For a given choice of $\ell$, this time is estimated as
\begin{equation}
t_{\rm log} = {\log N\over\gamma_\ell}.
\label{tlog}
\end{equation}
Physically, for $t > t_{\rm log}$, classical evolution produces
phase-space structures with areas less than $h = 1/N$, and so
quantum evolution is necessarily different.  We can define a
breakdown time in practice as the first value of $t$ for which
$\overline C_\ell(t) > \overline C_\ell(t-1)$.  The results are
compared with eq.~(\ref{tlog}) in fig.~(2) with good
agreement.

It is also interesting to look at off-diagonal correlation functions.
Let us consider the case $A(x)=(x|10)=P_1(q)=\sqrt{3}(1-2q)$ 
and $B(x)=(x|01)=P_1(p)=\sqrt{3}(1-2p)$. 
Classically, we can straightforwardly evaluate the single-step correlation 
$(01|\,{\cal U}|10)$; the result is $1-\sum_{a=1}^s w_a^3=1-u_2$.
Also, $(01|10)=0$ by orthogonality.  Since only the resonance $u_1$
can be involved in the time correlation of these distributions, 
and since the coefficient $c_1(t)$ can be at most linear in $t$ 
(because $u_1$ is two-fold degenerate), we must have
\begin{equation}
%(01|\,{\cal U}^t|10)=(1-u_2)\,t\,u_1^{t-1}
(01|\,{\cal U}^t|10)=(1-u_2) t u_1^{t-1}
\label{01u10}
\end{equation}
for $t\ge 0$.  (For any $t<0$, this correlation vanishes.) 

Quantum mechanically, it is also easy to evaluate the $t=1$ 
correlation function,
\begin{eqnarray}
{1\over N}\mathop{\rm Tr}P_1(p) U P_1(q) U^{-1}
&=& {1\over N} \sum_{jk} P_1(p_j)\langle p_j|U|q_k\rangle
                         P_1(q_k)\langle q_k|U^{-1}|p_j\rangle
\nonumber \\
&=& {1\over N} \sum_{jk} P_1(p_j) P_1(q_k) 
                         \Bigl|\langle p_j|U|q_k\rangle\Bigr|^2
\nonumber \\
&=& \sum_{a=1}^s {1\over w_a}
                 \int_{e_a}^{e_{a+1}} dp\,P_1(p) 
                 \int_{e_a}^{e_{a+1}} dq\,P_1(q)
                 + O(N^{-1}) 
\nonumber \\
&=& \sum_{a=1}^s {1\over e_{a+1}-e_a}\left[\sqrt{3}(e_{a+1}-  e_a 
                                                   -e_{a+1}^2+e_a^2)\right]^2
                 + O(N^{-1}) 
\nonumber \\
&=& 1-{\textstyle\sum_{a=1}^s} w_a^3 + O(N^{-1}).
\label{offd}
\end{eqnarray}
In the third line, we used eq.~(\ref{u}) and took the $N\to\infty$ limit.
To get the last line, we repeatedly used 
$\sum_{a=1}^s(e^\nu_{a+1}-e^\nu_a)=1$ for any exponent $\nu\ge 0$.
Eq.~(\ref{offd}) gives a simple example of how quantum expressions 
go over to their classical limits.

In fig.~(3), we show the quantum and classical expressions for this 
correlation function, for our two-strip example.  
To (slightly) improve the agreement, we also performed
an ensemble average over ten random choices of the four irrelevant parameters. 
(This can also be done for the diagonal correlation functions, with a similar
slight improvement.)

\section{Conclusions}
\label{6}

We have examined, both analytically and numerically, quantum time correlation
functions of the form $\mathop{\rm Tr}BU^t\!AU^{-t}$ in a generalized baker's
map, where $U$ is the quantum time evolution operator, and $A$ and $B$ are
smooth, $\hbar$ independent, functions of the position and momentum operators.
In the $\hbar\to0$ limit, these quantum correlation functions are governed
by the Ruelle resonances that govern the approach to ergodicity of the 
classical distributions $A$ and $B$ on phase space.  We see this very clearly
in our numerical results as long as the correlation function itself remains
larger than the predicted $O(\hbar)$ corrections.  These results also imply
that the statistical variance of the matrix elements of a smooth observable
(in the eigenbasis of $U$) is controlled by the Ruelle resonances.

\appendix
\section{Ruelle Resonances for the generalized Baker's Map}

The Ruelle resonances $u_\ell = e^{-\gamma_\ell}$ of the generalized 
baker's map are given by the zeros of the spectral determinant
\begin{equation}
d(z) = \det(1-z^{-1}{\cal U}).
\label{d1}
\end{equation}
We compute $d(z)$ following the analysis of Hasegawa and Saphir \cite{hs}
for the original baker's map.  We write
\begin{equation}
d(z) = \exp[ \mathop{\rm Tr} \ln(1-z^{-1}{\cal U})] 
     = \exp\biggl(-\sum_{t=1}^\infty{z^{-t}\over t}
       \mathop{\rm Tr}{\cal U}^t\biggr), 
\label{d2}
\end{equation}
which is valid for $|z|>1$.  The trace we need is given by
\begin{equation}
\mathop{\rm Tr}{\cal U}^t = \int dx\;\delta(x-{\cal M}^t(x))
                          = \int_0^1 dq_0 \int_0^1 
                            dp_0\;\delta(q_t-q_0)\,\delta(p_t-p_0),
\label{tr}
\end{equation}
where $q_t$ and $p_t$ are the values of $q$ and $p$ after $t$ 
iterations of the map.
To evaluate this trace, we must find the periodic orbits of period $t$.  
Each of these orbits is uniquely labelled by the set of strips in which we find
$q_0,q_1,\ldots,q_{t-1}$; the total number of these orbits is
$s^t$ (where $s$ is the number of strips).
The trace can then be evaluated as a sum over the periodic orbits,
\begin{eqnarray}
\mathop{\rm Tr}{\cal U}^t &=& \sum_{\rm p.o.} \,
    \left|\,\det\!\left({\partial x_t\over\partial x_0}-1\right)\right|^{\,-1}
\nonumber \\
          &=& \sum_{\rm p.o.}\,\Bigl|(\Lambda^{-1}-1)(\Lambda-1)\Bigr|^{\,-1}
\nonumber \\
          &=& \sum_{\rm p.o.}\sum_{\ell=0}^\infty (\ell+1)\Lambda^{\ell+1},
\label{tr2}
\end{eqnarray}
where we have defined
\begin{equation}
\Lambda \equiv \partial p_t/\partial p_0 
= (\partial q_t/\partial q_0)^{-1} 
= w(q_0)\ldots w(q_{t-1}),
\label{lambda}
\end{equation}
and $w(q_i)$ is the width of the strip in which we find $q_i$;
$\Lambda$ is related to the positive Lyapunov exponent $\lambda$ for the orbit
by $\Lambda = e^{-\lambda t}$.

Let $n_i$ be the number of times $q$ is in strip $i$ during a 
particular periodic orbit; $\sum_{i=1}^s n_i = t$.  Then, for that 
orbit,
$\Lambda = w_1^{n_1} \ldots w_s^{n_s}$,
and there are $t!/(n_1!\ldots n_s!)$ orbits with this value of $\Lambda$.
Thus we have
\begin{equation}
\mathop{\rm Tr}{\cal U}^t = \sum_{n_1=0}^t \ldots \sum_{n_s=0}^t 
          {t!\over n_1!\ldots n_s!} \,
          \delta_{n_1+\ldots+n_s,t} 
          \sum_{\ell=0}^\infty (\ell+1)\Lambda^{\ell+1}.
\label{tr4}
\end{equation}
We exchange the order of the summations to get
\begin{eqnarray}
\mathop{\rm Tr}{\cal U}^t &=& \sum_{\ell=0}^\infty (\ell+1)
            \sum_{n_1=0}^t \ldots \sum_{n_s=0}^t 
            {t!\over n_1!\ldots n_s!} \,
            \delta_{n_1+\ldots+n_s,t} \, 
            (w_1^{\ell+1})^{n_1} \ldots (w_s^{\ell+1})^{n_s}
\nonumber \\
        &=& \sum_{\ell=0}^\infty (\ell+1)
            (w_1^{\ell+1} + \ldots + w_s^{\ell+1})^t .
\label{tr5}
\end{eqnarray}
Inserting eq.~(\ref{tr5}) into eq.~(\ref{d2}) we get
\begin{equation}
d(z) = \exp\biggl(-\sum_{t=1}^\infty{z^{-t}\over t}
                \sum_{\ell=0}^\infty (\ell+1) u_\ell^t\biggr),
\label{d3}
\end{equation}
where we have defined
\begin{equation}
u_\ell \equiv w_1^{\ell+1} + \ldots + w_s^{\ell+1} .
\label{uell}
\end{equation}
Again exchanging the order of the summations we find
\begin{eqnarray}
d(z) &=& \exp\biggl(\,\sum_{\ell=0}^\infty 
(\ell+1)\ln(1-z^{-1}u_\ell)\biggr),
\nonumber \\
&=& \prod_{\ell=0}^\infty (1-z^{-1}u_\ell)^{\ell+1}.
\label{d4}
\end{eqnarray}
If we now analytically continue to $|z|<1$, we find zeroes of order $\ell+1$
at $z=u_\ell$ for $\ell=0,1,2,\ldots\;$.
This demonstrates that the Ruelle resonances are given by eq.~(\ref{uell}),
and that $u_\ell$ has degeneracy $d_\ell = \ell+1$.  Note that
$u_0 = 1$; the corresponding eigenfunction is the ergodic distribution, 
which is uniform over the unit square.

\begin{figure}
\label{fig1}
\epsfxsize=6in
\epsfbox{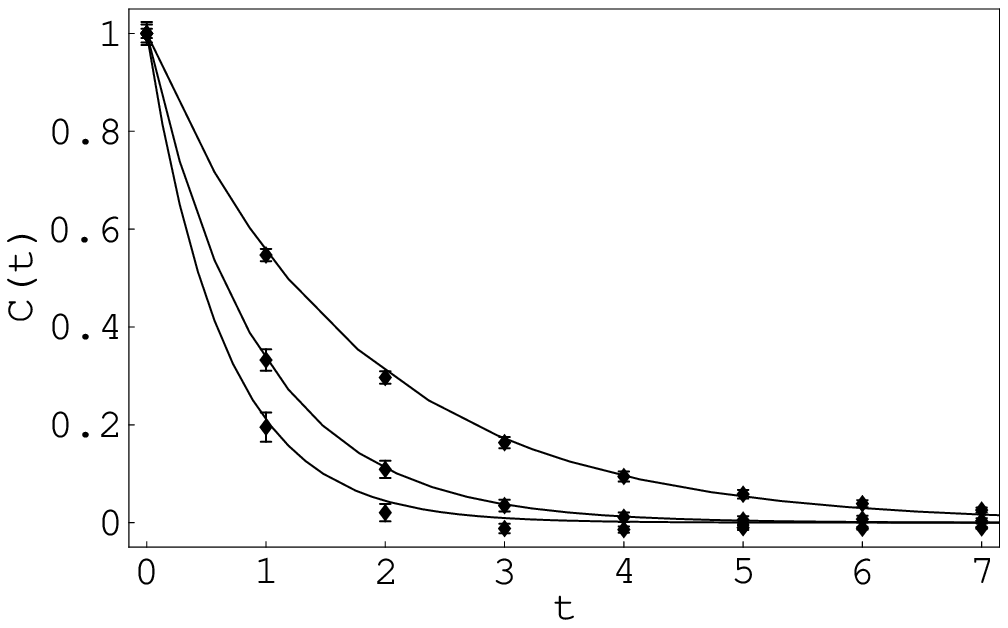}
\caption{Points show the correlation function
${\overline C}_\ell(t)
=(1/N)\mathop{\rm Tr}P_\ell(q)U^t P_\ell(q)U^{-t}$
for $\ell=1,2,3$, with $N=331$ and $w_1 = 110/331$;
error bars show the root-mean-square range of 
$C_{\ell,\alpha}(t)
=\langle\alpha|P_\ell(q)U^t P_\ell(q)U^{-t}|\alpha\rangle$.
Solid lines show the corresponding classical correlations $u_\ell^t$.}
\end{figure}

\begin{figure}
\label{fig2}
\epsfxsize=6in
\epsfbox{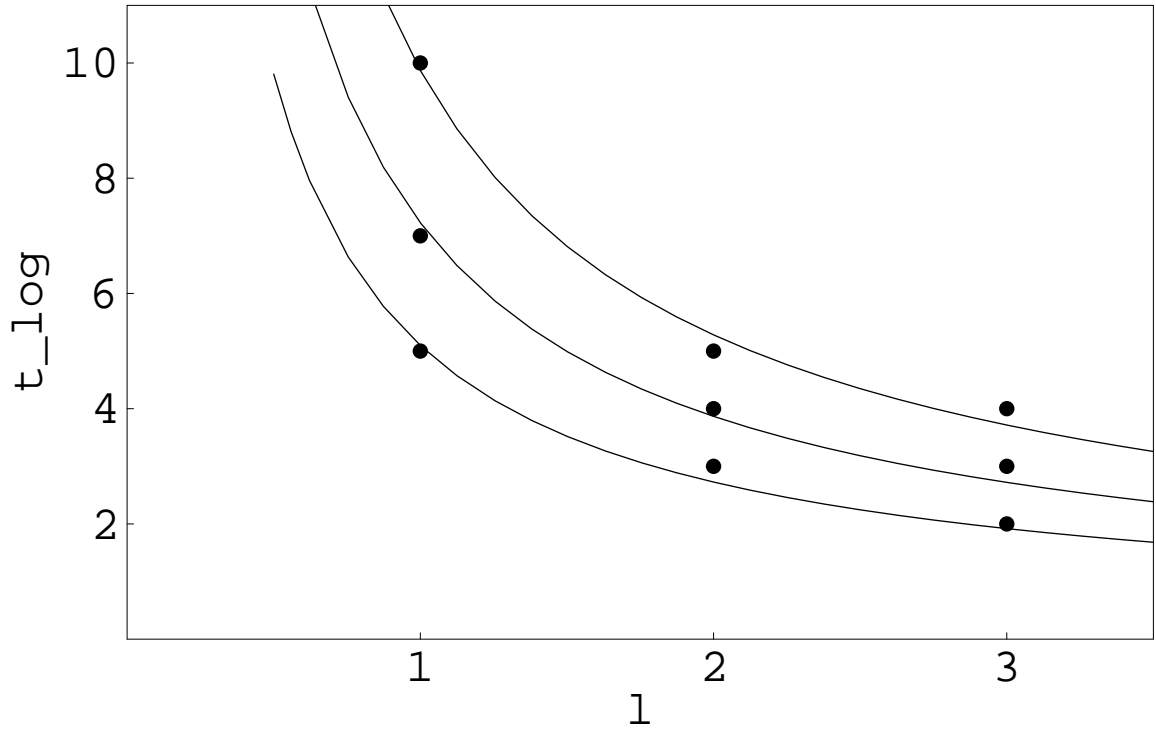}
\caption{Points show the break time $t_{\rm log}$, 
defined as the first value of $t$
for which the correlation function 
${\overline C}_\ell(t)$ increases, as a function of $\ell=1,2,3$,
for $N=21$, 72, and 331.  Solid lines show the
the prediction of eq.~(\ref{tlog}).}
\end{figure}

\begin{figure}
\label{fig3}
\epsfxsize=6in
\epsfbox{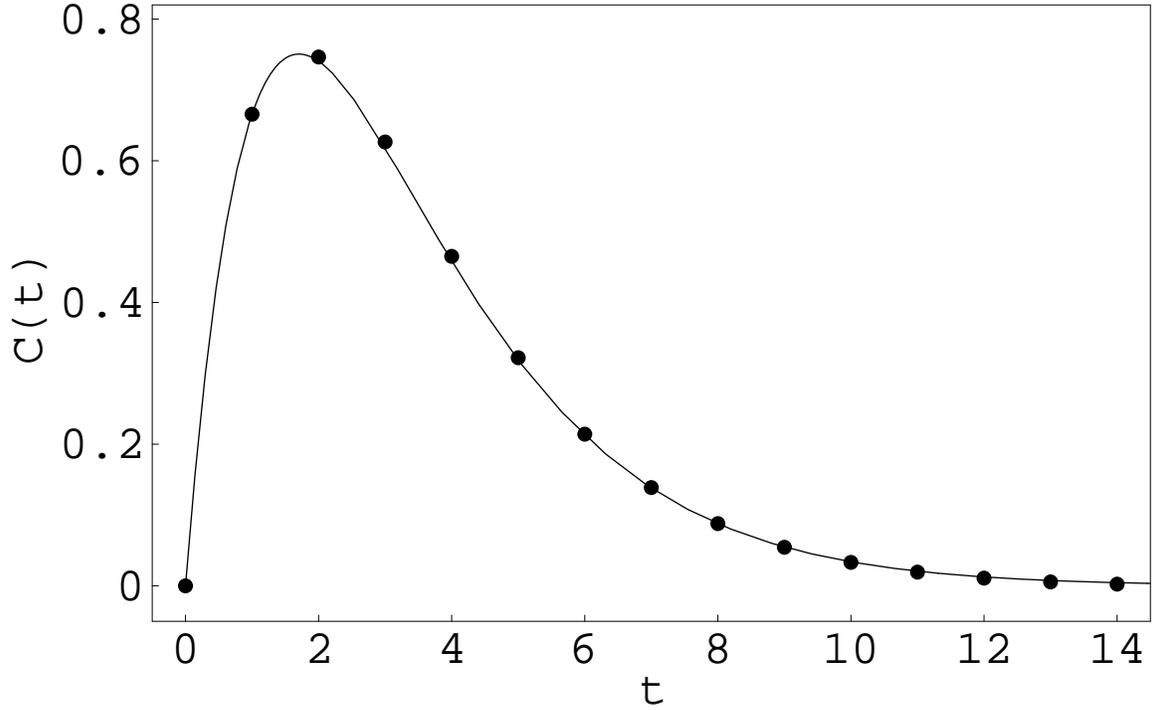}
\caption{Points show the off-diagonal correlation function
$(1/N)\mathop{\rm Tr}P_\ell(p)U^t P_\ell(q)U^{-t}$
for $N=331$ and $w_1 = 110/331$,
ensemble-averaged over ten random choices of the classically
irrelevant quantization parameters.
The solid line is the classical expectation $(1-u_2)tu_1^{t-1}$.}
\end{figure}

\end{document}